\documentclass[aps,reprint,amsmath,amssymb,floatfix,superscriptaddress]{revtex4-2}

\usepackage{soul}
\usepackage{graphicx}
\usepackage{dcolumn}
\usepackage{bm}
\usepackage{siunitx}
\DeclareSIUnit{\bits}{bits}
\DeclareSIUnit{\line}{l}
\usepackage[T1]{fontenc}
\usepackage{mathptmx}
\usepackage{makecell}
\usepackage{threeparttable}
\usepackage{xcolor}
\usepackage{nicefrac}
\usepackage{hyperref}
\usepackage{braket}
\usepackage{wasysym}
\usepackage{tikz}
\usepackage[english]{babel}
\usepackage{microtype}
\usepackage[version=4]{mhchem}

\usepackage{printlen}
\uselengthunit{in}

\definecolor{hotblue}{RGB}{46,48,146}
\hypersetup{
            colorlinks=true,
            linkcolor=hotblue,
            citecolor=hotblue,
            urlcolor=hotblue,
            }
\definecolor{blue}{RGB}{97,111,183}
\definecolor{orange}{RGB}{239,134,54}
\definecolor{green}{RGB}{76,167,96}
\definecolor{red}{RGB}{235,127,126}
\definecolor{yellow}{RGB}{235,235,100}

\begin{document}

\title{Calibration of systematic distortions in quantum emitter localization microscopy for deterministic nanophotonic fabrication}

\author{Chenxi Ma}
\thanks{Contributed equally to this work}
\author{Maximilian Heller}
\thanks{Contributed equally to this work}
\affiliation{Institute of Solid State Physics, Leibniz University Hannover, Appelstra\ss e~2, 30167~Hannover, Germany}

\author{Timon Handrup}
\author{Yiteng Zhang}
\affiliation{Institute of Solid State Physics, Leibniz University Hannover, Appelstra\ss e~2, 30167~Hannover, Germany}

\author{Tobias M. Krieger}
\author{Thomas Oberleitner}
\affiliation{Institute of Semiconductor and Solid State Physics, Johannes Kepler University Linz, Altenberger~Stra\ss e~69, 4040~Linz, Austria}

\author{Zenghui Jiang}
\author{Xian Zheng}
\author{Eddy P. Rugeramigabo}
\affiliation{Institute of Solid State Physics, Leibniz University Hannover, Appelstra\ss e~2, 30167~Hannover, Germany}

\author{Folke Dencker}
\affiliation{Institute of Micro Production Technology, Leibniz University Hannover, An~der~Universit\"at~2, 30823~Garbsen, Germany}

\author{Armando Rastelli}
\affiliation{Institute of Semiconductor and Solid State Physics, Johannes Kepler University Linz, Altenberger~Stra\ss e~69, 4040~Linz, Austria}

\author{Fei Ding}
\author{Michael Zopf}
\email{michael.zopf@fkp.uni-hannover.de}
\affiliation{Institute of Solid State Physics, Leibniz University Hannover, Appelstra\ss e~2, 30167~Hannover, Germany}
\affiliation{Laboratory of Nano and Quantum Engineering, Leibniz University Hannover, Schneiderberg~39, 30167~Hannover, Germany}

\begin{abstract}
{Quantum photonic technologies greatly benefit from quantum light emitters with high brightness, indistinguishability, and reliable polarization characteristics. Achieving optimal performance relies on the accurate localization of emitters and their deterministic integration into tailored photonic structures with nanometer-scale accuracy. Although marker-based photoluminescence imaging techniques can achieve statistical fitting uncertainties below \qty{10}{\nano\metre}, the ultimate integration yield is often limited by uncorrected systematic distortions in custom cryo-optical setups that compromise metrological accuracy. Here, we present an in situ calibration protocol that uses lithographically defined gold nanodisk arrays as references to calibrate optical distortions with a Zernike vector-field model. On held-out validation patterns beyond the calibration dataset, this correction reduces the residual systematic bias to \qty{5.3}{\nano\metre} with a 2D scatter of \qty{24.6}{\nano\metre} across the analyzed field of view. Furthermore, we demonstrate that applying this correction to the deterministic fabrication of circular mesa structures around semiconductor quantum dots reduces the variance in emission polarization by \qty{49}{\percent}, indicating improved registration accuracy. This calibration strategy offers a practical route to high-yield deterministic integration of quantum emitters into scalable quantum photonic circuits.}
\end{abstract}

\maketitle
\section{Introduction}

Photons serve as ideal carriers of quantum information due to their strong resilience to environmental decoherence, establishing quantum light sources as essential components for advanced photonic quantum technologies~\cite{Yin:2017,madsen_quantum_2022}. Among different types of quantum emitters, epitaxial semiconductor quantum dots (QDs) stand out as key sources for bright and highly indistinguishable single photons~\cite{wei_tailoring_2022,zhai_quantum_2022,ding_high-efficiency_2025}, entangled photon pairs~\cite{Basset2019,Zopf2019}, and photonic cluster states~\cite{cogan_deterministic_2023,coste_high-rate_2023}, with applications ranging from photonic quantum computing~\cite{maring_versatile_2024} to long-distance quantum communication~\cite{yang_high-rate_2024,strobel_telecom-wavelength_2025,laneve_quantum_2025,vajner_single-photon_2026}. In addition, their material platforms are compatible with mature semiconductor fabrication techniques, paving the way for scalable integration with photonic circuits~\cite{larocque_tunable_2024,wang_large-scale_2025}. 

Various photonic cavities and waveguides have been integrated with QDs to enhance light extraction, tailor spontaneous emission, and improve photon indistinguishability~\cite{somaschi_near-optimal_2016,liu_high_2018,liu_solid-state_2019,Uppu:2020,yang_tunable_2024,zhu_hybrid_2025,Wang2025Moire,mao_single-photon_2025}. However, scaling quantum photonic systems from single-emitter devices to multi-emitter architectures remains challenging because optimal integration requires precise spatial and spectral matching between individual emitters and target optical modes, together with preselection of emitters with favorable optical properties. These requirements are difficult to satisfy because high-quality epitaxial QDs are typically grown by self-assembly mechanisms, which lead to random spatial distribution and inhomogeneous wavelength broadening. Even displacements as small as tens of nanometers can already degrade the coupling efficiency and Purcell enhancement, and modify the polarization response of QD-nanostructure systems~\cite{Chu2020,Peniakov2024polarized,Gaur2026}. Therefore, deterministic integration strategies are essential: the position and spectroscopic properties of selected QDs must be obtained before fabricating photonic nanostructures.

To meet these demands, two strategies are generally employed: marker-free in situ lithography and marker-based optical positioning. Marker-free low-temperature optical lithography has achieved spatial accuracy within \qty{50}{\nano\metre} and spectral accuracy within \qty{0.5}{\nano\metre}~\cite{Dousse:2008,somaschi_near-optimal_2016}. However, since the fabrication pattern is defined by diffraction-limited optical exposure, this approach is less suited for complex, subwavelength nanostructures. Cathodoluminescence-integrated electron beam lithography (EBL) provides an alternative in situ route that bypasses the optical diffraction limit and has achieved QD-structure misalignments of \qty{34}{\nano\metre}~\cite{Gschrey:2015,gschrey_highly_2015}. In contrast, marker-based wide-field photoluminescence (PL) imaging adopts a different approach by decoupling emitter localization from subsequent lithographic patterning~\cite{sapienza_nanoscale_2015,Liu:2017,Pregnolato:2020,holewa_high-throughput_2024,rota_source_2024}. This separation enables flexible EBL fabrication while allowing emitters to be preselected via micro-PL for desired optical properties, such as emission wavelength, brightness, linewidth, and exciton fine-structure splitting. Building upon this conventional framework, recent advancements have incorporated hyperspectral imaging to further improve throughput and device yield by simultaneously extracting spectral information and spatial coordinates~\cite{liu_super-resolved_2024,buchinger_deterministic_2025}. This evolving wide-field PL imaging technique has become particularly attractive for the high-throughput selection and deterministic integration of high-quality emitters into scalable quantum photonic architectures.

\begin{figure*}[t]
\centering
\includegraphics[width=6.7in]{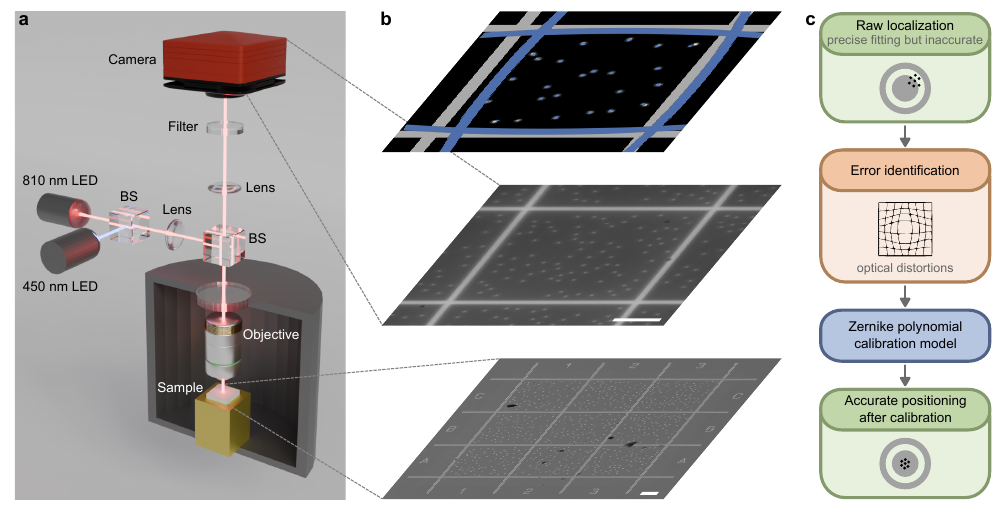}
\caption{Schematic of the optical setup and calibration workflow. (a) 3D illustration of the wide-field PL microscope setup for optical positioning. BS, beamsplitter. (b) From bottom to top: SEM image of a \numproduct{3 x 3} gold nanodisk calibration array; wide-field reflection image of the central marker field; and conceptual illustration comparing positions extracted from wide-field images (blue) with their nominal design positions (gray). Spatially varying residuals between the two coordinate sets are used to map the optical distortion field. Scale bars in the SEM and reflection images are \qty{10}{\micro\metre}. (c) Distortion calibration workflow.}
\label{fig:1}
\end{figure*}

Statistical localization uncertainties in marker-based optical positioning have been reduced to below \qty{10}{\nano\metre}. However, the ultimate integration yield is not only limited by fitting noise, but also by uncorrected systematic optical distortions. These errors are particularly relevant in custom cryogenic PL microscopes, where non-trivial, setup-specific distortions may differ between sample exchanges and measurement runs. Even carefully aligned imaging systems retain intrinsic geometric distortion and lateral chromatic aberration, as detailed by Copeland~\textit{et~al.}~\cite{Copeland2024}. This creates a critical gap between statistical localization precision and metrological positioning accuracy~\cite{madigawa2024}, motivating rigorous in situ calibration protocols for every measurement run and thus enabling real-time characterization of the distortion field of the imaging system to ensure accurate emitter localization.

In this work, we present an in situ metrological calibration framework for characterizing and correcting systematic optical distortions in cryogenic quantum emitter localization microscopy. Using lithographically defined gold nanodisk arrays as references, the measured field-dependent coordinate offsets in wide-field PL images are represented with a Zernike vector-field model. We validate this model using held-out nanodisk arrays and show that systematic position offsets are suppressed across the marker field. Applying this calibration routine to the deterministic fabrication of mesa structures around individual QDs results in reduced anisotropy in emission polarization and narrower device-to-device variation, indicating improved QD-mesa registration. This versatile method bridges the gap between localization precision and accuracy, providing a practical calibration strategy for improving the reliability of imaging-based quantum emitter localization in custom microscopy platforms.

\section{Optical setup characterization}

\begin{figure*}[htbp]
\centering
\includegraphics[width=6.5in]{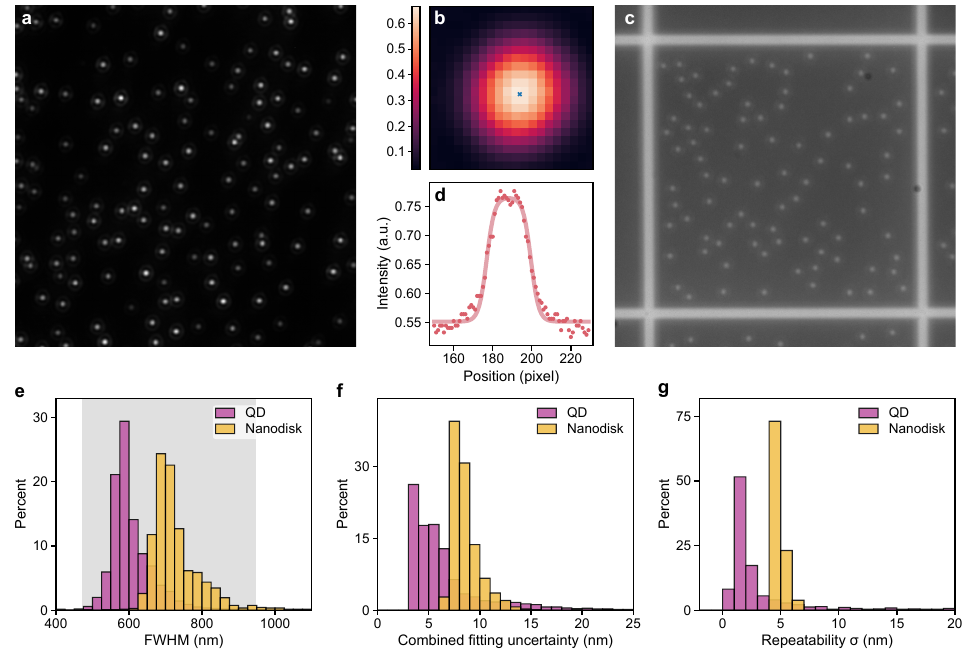}
\caption{Performance of the optical positioning setup. (a) Representative wide-field PL image of QD emission acquired through a \qty{785}{\nano\metre} narrow bandpass filter. (b) Magnified image of a 2D Gaussian fit for a representative QD. (c) Reflection image of the marker frame and nanodisks from the same field as (a), acquired through a \qty{785}{\nano\metre} narrow bandpass filter. The markers define an area of \qtyproduct[product-units=power]{40 x 40}{\micro\metre}. (d) Representative cross-section of a marker arm and the corresponding fit. (e) Distributions of the fitted FWHM values for QDs and nanodisks, with the shaded region spanning from the Gaussian approximation of the diffraction-limited PSF ($\mathrm{FWHM}_{\mathrm{diff}} = \mathrm{\qty{473}{\nano\metre}}$) to $2 \times \mathrm{FWHM}_{\mathrm{diff}}$. (f) Distributions of the localization fitting uncertainties for QDs and nanodisks. (g) Localization repeatability of QDs and nanodisks, quantified by the standard deviation of repeated localizations of the same object.}
\label{fig:2}
\end{figure*}

The optical positioning calibration technique is demonstrated using a sample with strain-free GaAs QDs grown by droplet etching and nanohole infilling~\cite{keil_solid-state_2017,huber_highly_2017}, which provides bright emission with a central wavelength around \qty{785}{\nano\metre}. After epitaxial growth, \qtyproduct[product-units = power]{40 x 40}{\micro\metre} alignment marker grids, consisting of a \qty{10}{\nano\metre} titanium adhesion layer and a \qty{50}{\nano\metre} gold layer, were patterned by EBL and metal lift-off. The sample was characterized at \qty{4}{\kelvin} in a closed-cycle cryostat equipped with a low-temperature objective with a numerical aperture (NA) of 0.82. The bichromatic wide-field imaging setup is mounted above the cryostat window, as illustrated in Fig.~\ref{fig:1}(a). The imaging scheme is designed to record QD emission and marker reflection within a common optical coordinate frame. In the horizontal illumination arm, a \qty{450}{\nano\metre} LED provides above-bandgap excitation for QDs, and an \qty{810}{\nano\metre} LED illuminates the alignment markers. Both beams merge at a beamsplitter, with the marker-illumination beam routed through the transmission port to facilitate coaxial alignment with the microscope. At the same time, \qty{90}{\percent} of the \qty{450}{\nano\metre} excitation power is retained in reflection, providing the excitation intensity required for rapid image acquisition. The LED beams are focused onto the back focal plane of the objective for better illumination uniformity. Image distortion is reduced by using a vertical collection path, which keeps the sample, objective, imaging lens, and camera close to a common optical axis. An air-spaced achromatic doublet lens with a focal length of \qty{100}{\milli\metre} images the sample onto a camera sensor with \numproduct{1920 x 1080} resolution and a pixel size of \qty{2.9}{\micro\metre}. The resulting field of view is \qty{\sim 76.5}{\micro\metre} wide in the sample plane, corresponding to an effective pixel size of \qty{\sim 70.8}{\nano\metre}. A \qty{785}{\nano\metre} narrow bandpass filter with a \qty{3}{\nano\metre} bandwidth is inserted into the collection path to reject the excitation light while transmitting QD emission. Owing to the broad spectrum of the \qty{810}{\nano\metre} LED, sufficient illumination for marker imaging still transmits through this filter.

We first evaluate the performance of the custom cryogenic microscope to establish the statistical precision of the localization pipeline. Figures~\ref{fig:2}(a,c) show a representative PL image of QDs and a reflection image of nanodisks (diameter \qty{\sim 740}{\nano\metre}) and alignment markers. The global coordinate frame is defined by localizing the alignment markers. Cross-sections extracted from individual pixel slices of each marker are fitted with a boxcar function convolved with a Gaussian approximation of the system point spread function (PSF), as shown in Fig.~\ref{fig:2}(d). To reduce errors caused by fabrication defects or nearby QD emission, slices with large fitting uncertainty or pronounced asymmetry are discarded. The marker centers are then obtained from the intersection of linear fits through the remaining slice centers, weighted by their fitting uncertainties. Single-QD emission spots are localized using 2D Gaussian fits, as shown in Fig.~\ref{fig:2}(b), where the fitted QD center is denoted by a cross. The full width at half maximum (FWHM) of the QD spots is centered at \qty{606.5}{\nano\metre}, as shown in Fig.~\ref{fig:2}(e). This value is slightly broader than the Gaussian approximation of the diffraction-limited microscope PSF, indicating proper focusing under experimental conditions~\cite{Zhang:07}. The combined QD-marker fitting uncertainty, obtained by propagating the covariance matrix of the 2D Gaussian QD fit together with the uncertainty of the marker-defined coordinate frame, has a mean of \qty{6.6}{\nano\metre}, as shown in Fig.~\ref{fig:2}(f). The corresponding independent uncertainties along the $x$ and $y$ directions are listed in Tab.~\ref{tab:errorcomparison}. The high-uncertainty tail is attributed primarily to the reduced signal-to-noise ratio of weakly detected QDs caused by spectral detuning from the bandpass filter. These results show that our imaging system provides sufficient image quality for reliable nanometer-scale localization under cryogenic conditions.

\begin{figure*}[htbp]
\centering
\includegraphics[width=7in]{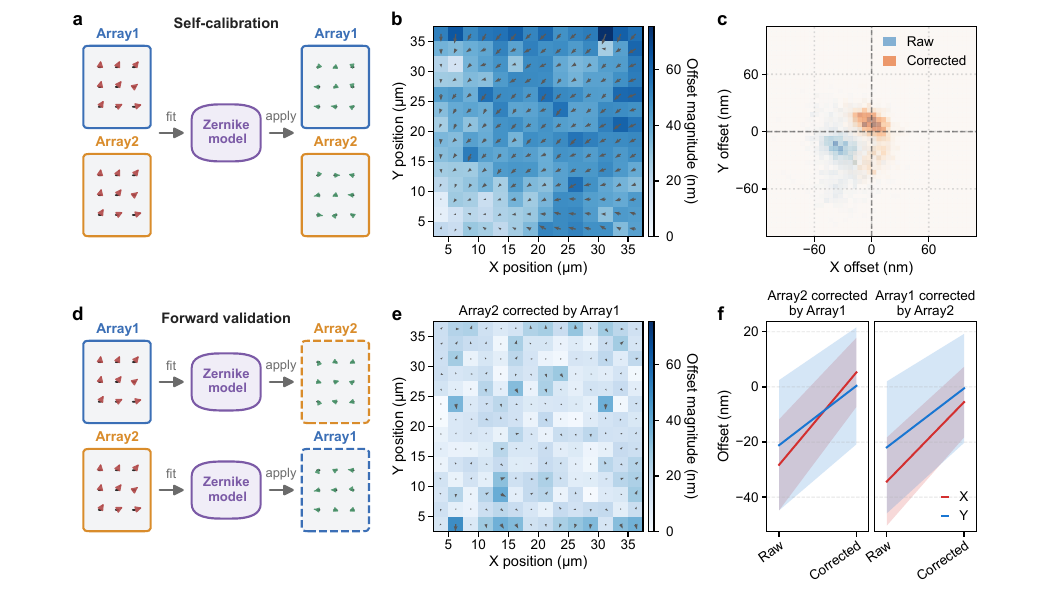}
\caption{Training and forward validation of the distortion-calibration model. (a) Schematic of the self-calibration process, in which a Zernike distortion model is fitted to the nanodisk offsets of both arrays. (b) Spatial map of the raw nanodisk offsets between optically extracted positions and design positions for both arrays. The arrows indicate the offset directions, and the color scale indicates the offset magnitude. (c) Overlaid 2D histogram of the nanodisk offsets before (blue) and after (orange) correction by the Zernike distortion model. (d) Schematic of the forward validation process. The model is trained on one array and then applied to the held-out array, after which the residual nanodisk offsets are evaluated. (e) Residual offset map of Array~2 after correction using the distortion model trained on Array~1. The same color scale as in (b) is used to visualize the reduction in offset magnitude. (f) Forward-validation performance for the two reciprocal train-test directions. Left: Array~2 corrected using the model trained on Array~1. Right: Array~1 corrected using the model trained on Array~2. The red and blue lines denote the $x$ and $y$ mean offsets, and shaded bands indicate the corresponding standard deviation.}
\label{fig:3}
\end{figure*}

Mechanical vibration from the cryostat cold head can blur wide-field PL images and degrade localization precision. To evaluate the mechanical stability of the cryostat and optical setup, we acquired 100 consecutive images of the same marker field and calculated the standard deviation of the fitted QD positions. Each image was acquired with an exposure time of \qty{4}{\second}, which is much longer than the vibration oscillation period; the measured positions therefore reflect the time-averaged stability of the imaging system. The resulting QD localization repeatability is \qty{3.9 \pm 6.2}{\nano\metre}, as shown in Fig.~\ref{fig:2}(g). This shows that our localization protocol is robust against setup vibration, image noise, and variations in the initial fitting parameters over repeated acquisitions.

\section{Distortion calibration}

\begin{table}[b]
    \caption{\label{tab:errorcomparison} Comparison of different localization errors}
    \begin{threeparttable}
    \begin{ruledtabular}
    \begin{tabular}{ccc}
    Type & \makecell[c]{Position error in $x$\\(nm)} & \makecell[c]{Position error in $y$\\(nm)}\\
    \colrule
    \makecell[c]{Fitting uncertainty\\(QD + marker)} & \num{4.9 \pm 3.1}  & \num{4.4 \pm 3.0} \\
    Optical distortion & \num{-31.3 \pm 16.5}  & \num{-21.5 \pm 23.7} \\
    EBL overlay alignment \cite{supplementary} & \num{11.0 \pm 13.6} & \num{-10.5 \pm 15.0} \\
    \end{tabular}
    \end{ruledtabular}
    \end{threeparttable}
\end{table}

A fundamental limitation is that low statistical localization uncertainty does not, by itself, guarantee the absolute positioning accuracy required for deterministic nanofabrication. Systematic optical distortions, schematically illustrated in the top panel of Fig.~\ref{fig:1}(b), can introduce position-dependent localization errors even when the fitting uncertainty is only a few nanometers. Because the true coordinates of self-assembled QDs are unknown, we use a lithographically defined calibration pattern whose design coordinates serve as known reference positions to quantify and correct these systematic errors.
The calibration pattern consists of 1600 gold nanodisks randomly distributed across two adjacent \numproduct{3 x 3} marker-grid arrays (Array~1 and Array~2). A scanning electron microscope (SEM) image of Array~2 is shown in the bottom panel of Fig.~\ref{fig:1}(b). Conventionally, in a periodic calibration pattern, the spacing between resolvable features is constrained by the diffraction-limited resolution of the microscope (\qty{\sim 479}{\nano\metre} in our setup). By contrast, our random distribution strategy probes a broader range of subfield coordinates; after registering multiple marker fields into a common coordinate frame, the resulting localization measurements can be superimposed to reconstruct the distortion field with an effectively higher spatial sampling density. The nanodisk patterns were imaged using the \qty{810}{\nano\metre} LED reflection and the \qty{785}{\nano\metre} narrow bandpass filter, and their reflection profiles were subsequently analyzed using the same 2D Gaussian fitting protocol applied to QDs. The nanodisk diameter was chosen to yield reflection profiles comparable in size to QD emission spots; however, the fitted nanodisk profiles show a larger mean FWHM, as shown in Fig.~\ref{fig:2}(e). This profile broadening arises from size enlargement due to fabrication imperfections and reduced contrast in the reflection images, which can also account for the slightly larger fitting uncertainty and poorer repeatability. Thus, the nanodisk-based experiments provide a conservative estimate of the distortion-correction performance expected for actual QD localization.

The optical distortions are quantified from the residual offset vectors between the detected and design positions of all nanodisks after transformation into the marker-defined coordinate frame, as mapped in Fig.~\ref{fig:3}(b). The arrows point from the detected positions to the design positions and therefore indicate the direction of the required correction, while their lengths and the cell colors represent the offset magnitudes. Each cell shows the average offset of all nanodisks falling within that spatial bin. A dominant pincushion-distortion pattern can be observed, with the distortion center located near the lower left corner of the marker field. As summarized in Tab.~\ref{tab:errorcomparison}, optical distortion is the largest error contribution in the uncorrected coordinate mapping, substantially exceeding the statistical fitting uncertainties. This necessitates digital distortion calibration to convert precise localization into accurate positioning.

Recent work has shown that standard computer vision calibration methods, such as Zhang's method, can yield high QD localization accuracy in deterministic nanofabrication workflows~\cite{Zhang:2020,buchinger_deterministic_2025}. Such general-purpose geometric models provide powerful numerical corrections, but they often describe the optical system through an effective projection matrix. For our custom cryogenic microscope, where field-dependent localization errors can arise from residual optical aberrations, alignment imperfections, and nonideal optical elements, we adopt a more physically interpretable representation. Specifically, we model the measured distortion field as a linear combination of Zernike polynomial terms up to the $n$-th Noll index ($n=10$), following the approach described by Copeland~\textit{et~al.}~\cite{copeland_subnanometer_2018}:
\begin{equation}
    \mathbf{x}_{\mathrm{corr}}
    =
    \mathbf{x}_{\mathrm{det}}
    +
    \sum_{i=1}^{n} Z_i(\rho,\varphi)\mathbf{p}_i
\label{eq:zernike}
\end{equation}
where $\mathbf{x}_{\mathrm{corr}}$ is the distortion-corrected position vector, $\mathbf{x}_{\mathrm{det}}$ is the detected position vector, $Z_i$ denotes the $i$-th Zernike polynomial in Noll indexing, and $\mathbf{p}_i$ is a coefficient vector describing the $x$ and $y$ correction components of this term. The polar coordinates $(\rho,\varphi)$ are defined with respect to the distortion center $\mathbf{p}_{\mathrm{c}}$.
For the calibration, the design coordinates are first transformed into the image coordinate system. We ensure the markers are in the same position during the acquisition of calibration images and QD images to minimize the error introduced by this transformation. Each transformed design coordinate is then paired with the nearest detected nanodisk position within a predefined distance threshold. The optimal model parameters $\mathbf{P} = \{\mathbf{p}_{\mathrm{c}}, \mathbf{p}_1,\ldots,\mathbf{p}_n\}$ are determined by minimizing residuals between the corrected detected positions and the transformed design coordinates:
\begin{equation}
\mathbf{P}_{\mathrm{opt}}
=
\operatorname*{arg\,min}_{\mathbf{P}}
\sum_{k=1}^{N}
\left\|
\mathbf{x}_{\mathrm{det},k}
+
\sum_{i=1}^{n}
Z_i(\rho_k,\varphi_k)\mathbf{p}_i
-
\tilde{\mathbf{x}}_{\mathrm{design},k}
\right\|_2^2
\label{eq:zernike_optimization}
\end{equation}
where $k$ indexes the $N$ calibration nanodisks. The minimization is performed using the Broyden–Fletcher–Goldfarb–Shanno algorithm implemented by SciPy~\cite{scipy2020}. Afterwards, the fitted model is evaluated by applying it to the detected coordinates. The corrected coordinates are transformed back into the marker-defined coordinate system and compared with the original lithographic design coordinates. After calibration, the residual positioning offsets are reduced to \qty{-0.2 \pm 12.3}{\nano\metre} along the $x$ direction and \qty{-0.8 \pm 19.1}{\nano\metre} along the $y$ direction, represented by the orange histogram in Fig.~\ref{fig:3}(c). Note that the off-centered distribution of the residual $y$ offset is associated with an EBL writing asymmetry, possibly due to dynamic beam-deflection effects. This imperfection can be removed from the calibration analysis if SEM-measured nanodisk positions are used as reference coordinates (see Supplementary Note 2). As a result, this calibration framework converts nanometer-scale localization precision into improved absolute positioning accuracy by explicitly correcting field-dependent optical distortion. In addition, the decomposition of different Zernike terms provides a physically interpretable description of the dominant distortion components, offering a useful diagnostic for future optimization of the imaging system.

\begin{figure*}[t!]
\centering
\includegraphics[width=7in]{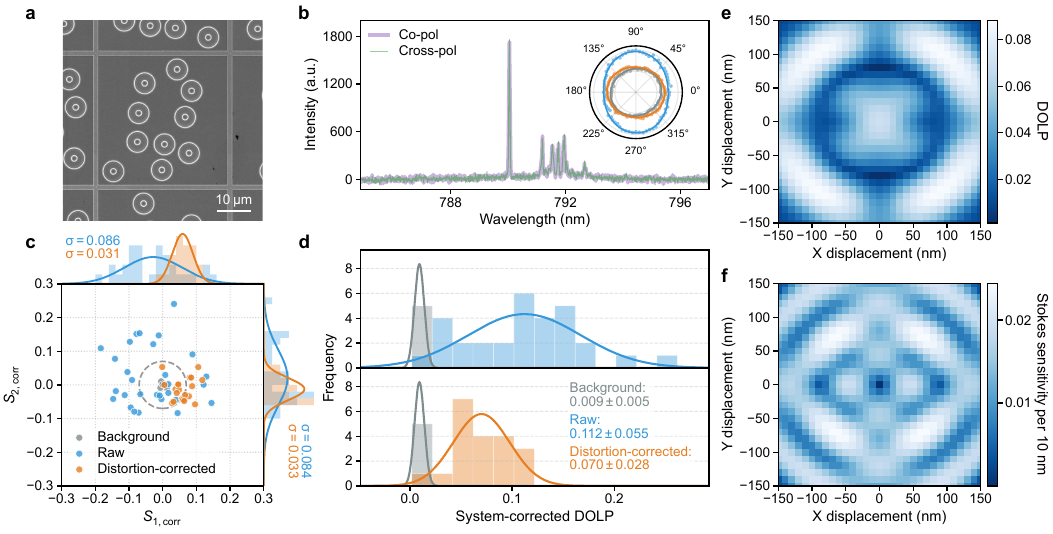}
\caption{Polarization response of deterministically fabricated mesas. (a) SEM image of a region inside the mesa array. (b) Background-subtracted co- and cross-polarized PL spectra from a distortion-corrected mesa, corresponding to the polarization angles of $I_{\max}$ and $I_{\min}$. The inset shows representative polar plots of the integrated PL intensity as a function of the linear polarization angle for a background QD (gray), a raw-positioned mesa (blue), and a distortion-corrected mesa (orange). (c) Scatter plot of the system-corrected Stokes components. The dashed circle indicates the DOLP expected for a centered emitter in an elliptical mesa with SEM-measured diameters. (d) Histograms of system-corrected DOLP for background QDs, raw-positioned mesas, and distortion-corrected mesas. (e) Simulated DOLP for different emitter displacements inside an elliptical mesa with SEM-measured diameters. (f) Simulated Stokes sensitivity, calculated from the gradient magnitude $\sqrt{\left|\nabla S_{1,\mathrm{sim}}\right|^2+\left|\nabla S_{2,\mathrm{sim}}\right|^2}$ and scaled to represent the expected local Stokes-vector change for a \qty{10}{\nano\metre} emitter displacement.}
\label{fig:4}
\end{figure*}

To assess the reliability and transferability of this calibration framework, we perform forward predictive validation. The Zernike distortion model is trained only on Array~1 by fitting the coordinate offsets, while Array~2 is excluded from this training procedure as a held-out validation field, as illustrated in Fig.~\ref{fig:3}(d). Before correction, the detected nanodisk positions in Array~2 exhibit raw offsets of \qty{-28.3 \pm 16.5}{\nano\metre} along the $x$ direction and \qty{-21.1 \pm 23.6}{\nano\metre} along the $y$ direction. These values are comparable to those obtained from the full dataset (Tab.~\ref{tab:errorcomparison}), indicating a consistent optical distortion across neighboring areas. We then apply the correction map derived from Array~1 to the detected coordinates in Array~2. As shown in Fig.~\ref{fig:3}(e) and the left panel of Fig.~\ref{fig:3}(f), the Zernike-based correction substantially suppresses the residual offsets in Array~2 to \qty{5.3 \pm 12.5}{\nano\metre} and \qty{0.4 \pm 21.2}{\nano\metre}, corresponding to a residual mean-vector bias of \qty{5.3}{\nano\metre} and a 2D scatter of \qty{24.6}{\nano\metre}. The outliers are attributed primarily to bins containing only a small number of nanodisks, which makes the local average less statistically robust. This limitation can be mitigated in future implementations by increasing the nanodisk density. The reciprocal validation, in which the distortion model trained on Array~2 is applied to Array~1, is shown in the right panel of Fig.~\ref{fig:3}(f). In both validation directions, the residual mean $x$ offset is not fully reduced to zero, indicating a small difference in the $x$ component of the distortion field between the two arrays. This observation suggests that calibration fields are better placed close to the target fabrication fields to minimize spatial variations in the distortion correction. Overall, this forward predictive validation demonstrates that the Zernike model captures reproducible, field-dependent optical distortion rather than merely overfitting the training data. These results also establish the nanodisk-based calibration model as a transferable correction strategy for optical localization of quantum emitters.

\section{Experimental validation}
Having validated the nanodisk-based distortion calibration model, we next apply it to QD localization and evaluate the resulting QD-device registration. In addition to marker reflection images, PL images of QDs were acquired using the \qty{450}{\nano\metre} LED and the \qty{785}{\nano\metre} narrow bandpass filter. After extracting the QD positions with the localization workflow described above, circular mesas were deterministically fabricated around selected QDs via EBL and chlorine-based dry etching, using either raw or distortion-corrected QD coordinates. A representative SEM image of fabricated mesas is shown in Fig.~\ref{fig:4}(a). Because the mesa polarization response is sensitive to the relative displacement between the QD and the mesa center, polarization-resolved micro-PL provides a probe of QD-mesa registration (see Supplementary Note 3). Under above-bandgap excitation, carrier relaxation largely removes memory of the excitation polarization. Therefore, the measured emission polarization is primarily governed by the QD-mesa system~\cite{Poem2011}. Figure~\ref{fig:4}(b) shows typical emission spectra from a QD embedded in a distortion-corrected mesa. The counts within the QD emission window are then integrated to obtain the PL intensity for each half-wave plate angle $\theta$ and fitted with $I(\theta) = I_{\min} + (I_{\max} - I_{\min}) \cos^2(2\theta - \phi)$, where $\phi$ is the principal polarization angle at which the maximum intensity occurs. As the fitting is performed independently for each emitter, our method of extracting only $I_{\max}$ and $I_{\min}$ is equivalent to the Stokes parameter measurement with fixed polarization bases, considering the circular polarization component is negligible according to simulations (see Supplementary Note 3). Therefore, the degree of linear polarization (DOLP) can be calculated as $\Pi = (I_{\max} - I_{\min})/(I_{\max} + I_{\min})$. We can also calculate the normalized linear Stokes vectors as $S_1 = \Pi \cos(2\phi)$ and $S_2 = \Pi \sin(2\phi)$. Note that these reconstructed components are defined in the analyzer basis rather than in the absolute H/V and D/A polarization bases. Background QDs without mesas were also measured as references, exhibiting a reproducible small DOLP with a preferred polarization angle, which we attribute to a system-induced polarization bias. The inset of Fig.~\ref{fig:4}(b) shows representative polarization-resolved PL intensities from a background QD, a raw-coordinate mesa, and a distortion-corrected mesa. To remove the system-induced bias, we use the average normalized Stokes vector of the background QDs as the system response function. The corrected linear polarization components can then be retrieved via vector subtraction:
\begin{equation}
\begin{aligned}
    S_{1,\mathrm{corr}} &= S_{1,\mathrm{mesa}} - \bar{S}_{1,\mathrm{ref}}\\
    S_{2,\mathrm{corr}} &= S_{2,\mathrm{mesa}} - \bar{S}_{2,\mathrm{ref}}
\end{aligned}
\label{eq:s1s2_subtract}
\end{equation}
The corrected DOLP is then reconstructed as $\Pi_{\mathrm{corr}}=\sqrt{S_{1,\mathrm{corr}}^2+S_{2,\mathrm{corr}}^2}$.

As shown in Fig.~\ref{fig:4}(c), the distortion-corrected mesas feature not only smaller $S_{1,\text{corr}}$ and $S_{2,\text{corr}}$ but also a more compact distribution, compared with raw-coordinate mesas. This trend further manifests itself in Fig.~\ref{fig:4}(d): the mean corrected DOLP decreases from 0.112 for raw-coordinate mesas to 0.070 for distortion-corrected mesas, corresponding to a reduction of approximately \qty{38}{\percent}. The device-to-device variation is also substantially reduced, with the standard deviation decreasing from 0.055 to 0.028, corresponding to an improvement of approximately \qty{49}{\percent}. These results indicate improved average QD-mesa centering and reduced placement dispersion after distortion correction.

The residual DOLP should not be interpreted as a direct measure of the emitter displacement, as it also contains contributions from structure-induced anisotropy. SEM measurements reveal a slight ellipticity of \qty{2.07 \pm 0.51}{\percent} of the fabricated mesas (see Supplementary Note 4). Finite-difference time-domain simulations using the measured mesa diameters (\qty{2.00}{\micro\metre} along the long axis and \qty{1.96}{\micro\metre} along the short axis) show that even a centered emitter can already produce an appreciable DOLP of approximately 0.070 (Fig.~\ref{fig:4}(e)), comparable to the mean experimental result. Furthermore, the collapse of the reconstructed Stokes vectors into a narrow distribution after distortion correction (Fig.~\ref{fig:4}(c)) can be interpreted in the following way: before correction, QDs are more likely to occupy a non-central position where Stokes-vector changes are highly sensitive to small position variations; after correction, QD positions are shifted towards the central area, where position variations lead to a narrower Stokes distribution, as can be deduced from Fig.~\ref{fig:4}(f). Therefore, the most plausible interpretation is that the distortion-corrected mesas contain near-centered QDs whose residual DOLPs are dominated by the structural ellipticity, with possible small contributions from residual displacement. Quantitative evaluation of the integration accuracy can be conducted by correlating cathodoluminescence mapping with SEM in future investigations.

\section{Conclusion}
In conclusion, we have developed a reference-based metrological framework that distinguishes statistical localization precision from absolute integration accuracy in deterministic quantum-emitter fabrication. Using lithographically defined gold nanodisks as a reference, we quantified the offset field between optically detected positions and design coordinates, and represented it with a Zernike vector-field model. This approach provides a physically interpretable calibration of field-dependent optical distortions. Applied to our cryogenic wide-field localization microscope, the calibration reduces the nanodisk mean offsets to nearly zero, with residual spreads approaching the statistical and fabrication-related limits. Forward validation on held-out nanodisk arrays demonstrates that the fitted Zernike model captures reproducible field-dependent distortions rather than merely overfitting the calibration data. Device-level validation using polarization-resolved PL measurements of QDs embedded in circular mesas further confirms improved registration accuracy: distortion correction yields both a reduced mean DOLP and a narrower device-to-device polarization distribution compared with raw-coordinate devices. The proposed method can be implemented in a broad range of emitter systems, custom microscopy setups, and nanofabrication workflows~\cite{Elshaari_hBN_2021}. Traceable and predictive positioning protocols of this type will be important for improving fabrication yield as quantum photonic devices evolve from isolated proof-of-concept structures toward larger-scale integrated circuits.

\section*{Methods}
\subsection{Sample preparation}
The QD sample was grown by molecular beam epitaxy (Riber C21T) on a GaAs wafer. The heterostructure consists of a \qty{200}{\nano\metre} GaAs buffer layer, a \qty{400}{\nano\metre} \ce{Al_{0.75}Ga_{0.25}As} sacrificial layer, and a \qty{130}{\nano\metre} \ce{Al_{0.23}Ga_{0.77}As} barrier layer, where strain-free GaAs QDs were formed at the center by droplet etching and nanohole infilling. The barrier layer was capped on both sides by \qty{10}{\nano\metre} GaAs layers to suppress oxidation. The alignment markers were fabricated using standard EBL (Raith PIONEER Two) and metal lift-off processes. The mesa structures were patterned by EBL with a positive electron-beam resist (Allresist AR-P 6200.13), and subsequently etched into the QD sample using inductively coupled Ar/\ce{Cl_{2}}/\ce{BCl_{3}} plasma.
\subsection{Polarization-resolved PL}
The PL measurements were performed at cryogenic temperature using a closed-cycle cryostat (Montana Instruments). QDs were excited above-bandgap using a \qty{635}{\nano\metre} continuous-wave laser, which was rejected from the detection path by a \qty{750}{\nano\metre} longpass filter. The emission polarization was analyzed by a rotating half-wave plate followed by a fixed linear polarizer, placed directly after the objective and beamsplitter. The QD emission was then guided into a spectrometer (Princeton Instruments) equipped with a \qty[per-mode=symbol]{1800}{\line\per\milli\meter} grating for high spectral resolution and a \qty{500}{\micro\metre} wide slit to reduce sensitivity to the exact collection position~\cite{Laneve2025Wavevector}. For each measurement, the excitation power was kept below saturation. The spectral background was estimated from the average intensity in the first and last \qty{1.5}{\nano\metre} ranges and subtracted from the spectra.
\subsection{Numerical simulation}
Simulations were performed using Ansys Lumerical FDTD. The QD was modeled as an incoherent and unpolarized emitter by averaging the far-field intensities generated by two orthogonal in-plane electric dipoles oriented along the $x$ and $y$ directions. The dipole was placed at \qty{75}{\nano\metre} below the sample surface, same as the actual QD heterostructure. The mesa geometry was defined according to the measured device dimensions, with a long diameter of approximately \qty{2.00}{\micro\metre}, a short diameter of \qty{1.96}{\micro\metre}, and a height (etch depth) of approximately \qty{600}{\nano\metre}. The far-field intensity was integrated over the collection cone corresponding to the experimental objective (Mitutoyo LCD Plan Apo NIR HR 100X, NA = 0.7).

\section*{Author Contributions}
Y.Z. and E.P.R. grew the QD samples. C.M. fabricated the sample with the assistance of X.Z. and F.De. C.M. carried out optical measurements with the support of M.H. and Z.J. M.H. wrote the image analysis algorithm with contributions from T.H. T.M.K., T.O., and A.R. performed complementary optical measurements at their facility. C.M. and M.H. analyzed and interpreted the results together. C.M. prepared the first manuscript draft and revised it with M.H. and M.Z., with input from all other co-authors. M.Z. conceived and supervised the project together with F.D.

\section*{Acknowledgments}
We thank Tobias Huber-Loyola, Quirin Buchinger, and Qiushuang Lian for fruitful discussions. We also thank Rolf J. Haug, Ronny H\"uther, J\"urgen Becker, and Lukas~Steinhoff for their technical support.

\section*{Funding}
The authors gratefully acknowledge the funding from the German Federal Ministry of Research, Technology and Space (BMFTR) within the projects SemIQON (13N16291) and QVLS-iLabs: Dip-QT (03ZU1209DD), the project CLICS funded by the funding program zukunft.niedersachsen of the Lower Saxony Ministry of Science and Culture and the Volkswagen Foundation, the European Research Council (MiNet – No. GA101043851), the Deutsche Forschungsgemeinschaft (DFG, German Research Foundation) under Germany's Excellence Strategy (EXC-2123) QuantumFrontiers (390837967). A.R. acknowledges support of the Austrian Science Fund (FWF) [10.55776/COE1, 10.55776/PIN4389523, 10.55776/FG5] and the QuantERA program via the project MEEDGARD (FFG Grant No. 906046). T.M.K. acknowledges support of the Linz Institute of Technology (LIT), via the LIT project PEPSI (LIT-2025-14-YOU-121), that received funding from the State of Upper Austria and the Austrian Federal Ministry of Women, Science and Research. Y.Z. acknowledges the China Scholarship Council (CSC201908370225).

\section*{Disclosures}
The authors declare no conflicts of interest.

\nocite{}
\bibliography{reference}
\end{document}